\documentstyle[hx37_latex]{article}
\begin{document}
\title{THE UNIVERSE AT Z $>$ 5:  WHEN AND HOW DID THE `DARK AGE' END?}

\author{MARTIN J. REES}

\address{Institute of Astronomy, 
Madingley Road, Cambridge, CB3 OHA, UK}

\maketitle\abstracts{
This paper considers how the first subgalactic structures
  produced the UV radiation that ionized the intergalactic medium
  before $z = 5$, and the `feedback' effects of the UV radiation on
  structure formation. The relevance of pregalactic activity to heavy
  element production and the origin of magnetic fields is briefly
  addressed.}

\section{Introduction} 
  When the primordial radiation cooled below a few thousand
degrees, it shifted into the infrared. The universe then entered a
dark age, which continued until the first bound structures formed,
releasing gravitational or nuclear energy that lit up the universe
again.  How long did the `dark age' last? We know that at least some
galaxies and quasars had already formed by a billion years. But how
much earlier did structures form, and what were they like?

 The density of quasars and large galaxies
thins out at observed redshifts, but subgalactic structures may
exist even at redshifts exceeding 10.  I shall discuss the effects of
the earliest stars and supernovae -- production of UV radiation,
reheating of the IGM, and the production of the first heavy elements
-- and the implications for observations at ultra-high redshifts.

\section{Clustering in hierarchical models}
     
I will focus on the cold dark matter (CDM) model. But this is just a
`template' for some more general deductions, which essentially apply
to any `bottom up' model for structure formation.  There is no minimum
scale for the aggregation, under gravity, of cold non-baryonic
matter.$^{1-3}$ However the baryons constitute a gas whose pressure opposes
condensation on very small scales. The gas therefore does not `feel'
the very smallest condensations. The baryonic Jeans mass is
\begin{equation}
 M_J = 3 \times 10^5 \left({1 + z \over 10}\right)^{-{3 \over 2}} 
\left({T_g \over 500 {\rm K}}\right)^{3 \over 2} 
{\Omega b \over \Omega} M_\odot
\end{equation}
  On scales larger than this, baryons can condense into bound systems,
along with the dark matter.$^{1-3}$  During the `dark age' the gas became even
cooler than the microwave background: if it had cooled adiabatically,
with no heat input since recombination, its temperature $T_g$ would, at 
$z
= 10$, have been below 5 K.  The smallest bound structures, with mass 
$\sim M_J$, would have virialised at temperature of a few times larger than 
$T_g$. Larger masses would virialise at temperatures higher by a further
factor $(M/M_J)^{2/3}$. This virial temperature would be reached not solely
by adiabatic compression, but also because of a shock: it is unlikely
that the gas could contract by more than a factor of 2 in radius
before being shocked.
   
These virialised systems would, however, have a dull existence as
stable clouds unless they could lose energy and deflate due to atomic
or molecular radiative processes --- clouds that couldn't cool would
simply remain in equilibrium, being later incorporated in a larger
scale of structure as the hierarchy builds up.  On the other hand,
clouds that can cool will deflate, even  go
into free-fall collapse, and (perhaps after a disc phase) fragment
into smaller pieces.

    Three `cooling regimes' are relevant during successive phases of the
cosmogonic process, each being associated with a characteristic temperature.

    1. For a H-He plasma the only low-temperature ($< 10^3$ K) cooling
comes from molecular hydrogen. This cuts off below a few hundred
degrees; above that temperature it allows contraction within the
cosmic expansion timescale. The H$_2$ fraction is never high, and it is
in any case not a very efficient coolant
(eg Fig 1 of Tegmark {\it et al.}$^1$) but molecular cooling almost certainly
played a role in forming the very first objects that lit up the
universe.

   2. If H$_2$ is prevented from forming,  
then a H-He mixture behaves adiabatically unless $T$ is as
high as 8-10 thousand degrees, when excitation of Lyman alpha by the
Maxwellian tail of the electrons provides efficient cooling whose rate
rises steeply with temperature;  gas in this regime contracts almost
isothermally.

    3. The UV from early stars will photoionize some (and eventually
almost all) of the diffuse gas. When this happens, the HI fraction is
suppressed to a very low level, so there is is no cooling by
collisional excitation of Lyman lines; moreover the energy radiated
when a recombination occurs is quickly cancelled by the energy input
from a photoionization, so the only net cooling is via bremsstrahlung.
The cooling is, in effect, then reduced by a factor of $\sim 100$ 
(see, for
instance, ref 4). The minimum temperature (below which there is a net
heating from the UV) depends on the UV spectrum, and on whether  He
is doubly ionized: it is in the range 20-40 thousand degrees.  
  
\section{The role of molecular hydrogen, and the UV feedback} 

The role of molecular cooling at early cosmic epochs has been
   considered by many authors, dating back to the 1960s; recent
   discussions are due to Tegmark et al.$^1$  and Haiman et al.$^5$ 
This
   process allows clouds to contract if their temperature exceeds 
$\sim 500$
   K. The exact efficiency depends on the density , and therefore on
   the redshift when the first collapse occurs.

  But even at high redshifts, H$_2$ cooling would be quenched if there
were a UV background able to dissociate the molecules as fast as they
form. Photons of $h\nu > 11.18$ eV can photodissociate H$_2$, as first
calculated by Stecher and Williams.$^6$ These photons can penetrate a
high column density of HI and destroy molecules in virialised and
collapsing clouds, even when they are far less intense than the
background needed to fully ionize the medium.$^5$ 
 {\it   Only a small fraction of the UV that ionized the IGM can 
therefore have been
produced in systems where star formation was triggered by molecular
cooling. Most must have formed in systems large enough to have been
able to cool by atomic line effects.}

  There is then a further transition when the medium becomes
 completely ionized: the UV background gets a boost, because the
 contributions from remote regions (which dominate in Olbers-type
 integrals) are less severely attenuated. This means that it can maintain high
 ionization of a cloud until it has either collapsed to an overdensity
 exceeding the IGM ratio of ions to neutrals, or until it becomes
 self-shielding (which happens at more modest overdensities for large
 clouds). Until that happens the cooling rate will be reduced by the
 elimination of the (otherwise
 dominant) `line' contribution to the cooling.

  When this third phase is reached, the thermal properties of the
uncollapsed gas will resemble those of the structures responsible for
the observed Lyman-forest lines in high-z quasars spectra-- these are
mainly filaments, draining into virialised systems. Such systems have
velocity dispersions of $\sim 50$ km/sec, and will turn into
galaxies of the kind whose descendents are still recognisable.

\section{The first stars: some uncertainties}

The three uncertainties here are:

(i) What is the IMF of the first stellar population? The high-mass
stars are the ones that provide efficient (and relatively prompt)
feedback. It plainly makes a big difference whether these are the
dominant type of stars, or whether the initial IMF rises steeply
towards low masses, so that very many faint stars form before there is
a significant feedback.

(ii) The influence of the early stars depends on where their energy
is deposited. The UV radiation could, for instance, be mainly
absorbed in the gas immediately surrounding the first stars, so that
it exerts no feedback on the condensation of further clumps -- the
total number of massive stars needed to build up the UV background,
and the concomitant contamination by heavy elements, would then be
greater.

(iii) Quite apart from the uncertainty in the IMF, it is also unclear
what fraction of the baryons that fall into a clump would actually be
incorporated into stars before being re-ejected. The retained fraction
depends on the virial velocity: gas more readily escapes from shallow
potential wells.  Ejection is even easier in potential wells so
shallow that they cannot confine gas at the photoionization
temperature.

  All these three uncertainties would, for a given fluctuation
spectrum, affect the redshift at which molecules were destroyed, and
the (smaller) redshift at which full ionization occurred.

\section{Heavy elements, magnetic fields, and the oldest stars}

  If the main UV source is stars, there is inevitably an associated
build-up of heavy elements. (In more radical pictures where black
holes are involved in the early energy input, this inference doesn't
hold, because the energy supply could be gravitational rather than
nuclear). The question then arises of how this processed gas would be
distributed. Would it be confined in the virialised systems, or could
it spread through the entire IGM?
 
    The ubiquity of carbon features in intermediate and high ($N >
3.10^{14}$ cm$^{-2}$) column density systems (reported by other
speakers) implies that heavy elements are broadly enough dispersed to
have a large covering factor. These absorption systems may be
associated with the subgalactic $(\sim 10^9 M_\odot)$ sites of star
formation. The nucleosynthesis sites cannot therefore be too sparse if
these elements are, within the time available, to diffuse enough so
that they are encountered somewhere along every line of sight through
a typical high-column-density cloud. The absorption line data tell us
the {\it mean} abundance through the relevant cloud. They are
compatible with 99 percent of the material being entirely unprocessed,
and the heavy elements being restricted to 1 percent of the material
-- the early heavy elements need not be thoroughly mixed, but they
must have spread sufficiently to have a large `covering factor' in the
intermediate- and high-N clouds.

     The first stars are important for another reason: they may
generate the first cosmic magnetic fields. Moreover, mass loss (via
winds or supernovae permeated by magnetic flux) would disperse
magnetic flux along with the heavy elements. This flux, stretched and
sheared by bulk motions, can be the `seed' for the later amplification
processes that generate the larger-scale fields pervading disc
galaxies.
 
   The efficiency of early mixing is important for the interpretation
of stars in our own galaxy that have ultra-low metallicity -- lower
than the mean metallicity that would have been generated in
association with the UV background at $z > 5$.  If the heavy elements
were efficiently mixed, then these stars would themselves need to have
formed before galaxies were assembled. To a first approximation they
would thereafter cluster non-dissipatively; they would therefore be
distributed in halos (including the halo of our own Galaxy) like the
dark matter itself. More careful estimates slightly weaken this
inference, This is because the subgalaxies would tend, during the
subsequent mergers, to sink via dynamical friction towards the centres
of the merged systems. There would nevertheless be a tendency for the
most extreme metal-poor stars to have a more extended distribution in
our Galactic Halo, and to have a bigger spread of motions.

The number of such stars depends on the early IMF. If this were flatter,
there would be fewer low-mass stars formed concurrently with those
that produced the UV background. If, on the other hand, the IMF were
initially steeper, there could in principle be a lot of very low mass
(macho) objects produced at high redshift. These could be distributed
like the dark matter. They could provide a few percent of the halo if
$\Omega$ were 1; a larger proportion in a low-density universe.

\section{Summary}
  
  There are thus three stages in the build-up of hierarchical
structure, characterised by different masses and virial
temperatures. They  occur at three successive  epochs -- however, the
demarcation is unlikely to be sharp because the range of amplitudes
(for gaussian fluctuations) translates into a broad spread of
turnaround times for a given mass scale.

    These general conclusion are relevant to any model where the
initial fluctuations have amplitudes decreasing with scale, so that
cosmic structures form `bottom up'. Such models differ, of course, in
the epoch at which `first light' would have occurred. In PIB models,
this may be at $z> 100$; for CDM it is in the range 10-20; for `mixed
dark matter' models the first structures may form still more
recently. Molecular cooling tends to be more efficient at high
densities, and therefore at large redshifts; but in all cases it
determines the scale of the first objects that condense out and
contribute the first injection of heat into the universe.

     The amount of background UV generated per solar-mass of material
in these first objects is very uncertain -- it depends on the
efficiency of star formation,on whether the IMF favours massive stars
(or even supermassive objects or black holes), and on how much of the
UV is `soaked up' by dense gas within the bound objects themselves.
But irrespective of all these uncertainties, the UV background exerts
an important feedback on the cosmogonic process, by quenching H$_2$
cooling, long before  photoionizing  the
entire IGM.

   We therefore draw the robust conclusion that the IGM remained
predominantly neutral until a sufficient number of objects above $\sim
10^9((1 + z)/10)^{-3/2}M_\odot$ had gone non-linear.  Such systems 
have virial temperatures above 10,000 K -- hot
enough for HI line emission to permit very efficient cooling. Most of
the O-B stars (or accreting black holes) that photoionized the IGM had
to form in systems at least as large as this.

     Formation of such systems would have continued unimpeded until
the universe became, in effect, an HII region. This
must have happened before $z=5$. The only net cooling of a fully
photoionized gas comes from bremsstrahlung, which is less effective
than the collisionally-excited line emission from gas that is only
partly ionized The completion of photionization may therefore signal
another pause in the cosmogonic process,$^{7-8}$ associated with a further
increase in the minimum scale that can collapse, and in the efficiency
of cooling.

    By the epoch $z = 5$, some structures (albeit perhaps only
exceptional ones) must have attained galactic scales. Massive black
holes (manifested as quasars) accumulate in the deeper potential wells
of these larger systems (see, for instance, ref 9); quasars may
dominate the UV background at $z < 4$.

\section*{Acknowledgments}
I am grateful to Hugh Couchman with whom I first studied this subject
more than 10 years ago. I also thank my recent collaborators,
especially Zoltan Haiman, Avi Loeb and Max Tegmark.

\end{document}